\documentclass[9pt,twocolumn,english]{article}
\usepackage{amssymb}
\usepackage{amsmath}
\usepackage{amsbsy}
\usepackage{wasysym}
\usepackage{marvosym}
\usepackage{natbib}
\usepackage{url}
\usepackage{graphicx}
\usepackage{fancyvrb}

\title{Fast ordered sampling of DNA sequence variants}
\author{Anthony J. Greenberg \footnote{Bayesic Research, Ithaca, NY, USA; tonyg@bayesicresearch.org}}
\date{}

\begin{document}

\maketitle

\begin{abstract}
Explosive growth in the amount of genomic data is matched by increasing power of consumer-grade computers. Even applications that require powerful servers can be quickly tested on desktop or laptop machines if we can generate representative samples from large data sets. I describe a fast and memory-efficient implementation of an on-line sampling method developed for tape drives 30 years ago. Focusing on genotype files, I test the performance of this technique on modern solid-state and spinning hard drives, and show that it performs well compared to a simple sampling scheme. I illustrate its utility by developing a  method to quickly estimate genome-wide patterns of linkage disequilibrium (LD) decay with distance. I provide open-source software that samples loci from several variant format files, a separate program that performs LD decay estimates, and a C++ library that lets developers incorporate these methods into their own projects.
\end{abstract}
\section*{Introduction}
Growth in the amount of genomic data available is matched by increasing power and storage space of consumer-grade computers. Using such low-cost systems to perform genomis analyses can speed development cycles and empower users operating under economic constraints. The range of such analyses can be extended with light-weight software tools that carefully manage system resources.

Collections of single-nucleotide polymorphisms (SNPs) and copy number variants (CNVs) genotyped in groups of individuals are fundamental to numerous applications. These data sets are stored in a variety of formats \citep{hapmap03,purcell07a,danecek11a} and often contain millions of variants genotyped in thousands of individuals. It is often desirable to create random subsets of such large polymorphism tables. For example, relatively small samples can be used to quickly test software pipelines. In addition, when using genome-wide SNPs to predict phenotypes of individuals using genome selection methods, it is often important to learn the minimal marker set that achieves good accuracy \citep{spindel15a}. Finally, repeated creation of data subsets is a variant of the jack-knife procedure \citep{efron79} and can be used to construct empirical distributions of genome-wide statistics.

To be useful for a wide range of applications, any sampling scheme must meet several criteria. The subset generated must be in the same order as in the original data set. Variants must be sampled without replacement, each locus has to be picked with the same probability, and the size of the resulting data set must always reflect the value required by the user. The time required to sample variants must grow at most linearly with the number of polymorphisms in the sub-sample. Furthermore, because the original data set may be very large, the time required to pick loci must be as insensitive as possible to the overall size of the data set. In addition, since my aim is to empower researchers with limited access to powerful hardware, the implementation should minimize the use of system resources, particularly avoiding reading large files into memory. Surprisingly, after an extensive search I was unable to find existing software that performs according to these criteria. However, the general problem of ordered on-line sampling of records from files was solved 30 years ago \citep{vitter84a,vitter87a}. Unfortunately, this work is relatively unknown with limited application in computer system management and sampling of business data streams. No genetics papers appear to reference Vitter's articles.

I implemented a version of Vitter's algorithm \citep{vitter87a} that samples loci from a variety of variant file formats while minimizing system resource use. I examine the method's performance compared to a simple sampling scheme and provide an example application to estimate genome-wide patterns of linkage disequilibrium. I also provide a library that allows developers to incorporate these methods into their software. All source code, data, and analysis methods are openly available.

\section*{Methods}
\subsection*{Sampling scheme}
Vitter's main insight was to derive a scheme that samples the number of records to skip given how many remain to be picked and the number left in the file \citep{vitter84a,vitter87a}. There are additional speed-ups available if one is willing to store the index values in an array, but I opted to save memory instead and save the sampled records to the output file right away. Preliminary tests suggested that file I/O time dominates random number generation even on a machine with a solid state drive (SSD, results not shown), so the increase in sampling speed would not be noticeable.

Other than the above deviation, I implemented Vitter's method D as described in the Appendix A, algorithm A2 in \citet{vitter87a}. The implementation uses a hardware random number generator (RNG) (\url{https://software.intel.com/en-us/articles/the-drng-library-and-manual}). If not supported, the software substitutes the 64-bit Mersenne Twister \citep{matsumoto98a} seeded with the processor's time stamp counter. The decision is made automatically at run time and does not involve user input.

\subsection*{Included software}
This report describes three pieces of software: a C++ library \texttt{libsampFiles} and two stand-alone programs: \texttt{sampleSNPs} produces ordered samples from variant files and \texttt{sampleLD} uses locus samples to calculate distributions of linkage disequilibrium statistics. All software is released under the BSD three-part license. The whole set of programs can be trivially compiled using the included Makefile, with no external dependencies required. Compilation was tested on Mac OS with llvm/clang and on RedHat Linux using the GNU compiler collection.

\paragraph*{C++ class library.}
The \texttt{libsampFiles} library allows users to easily include in their own software support for sampling loci from most commonly used file formats (.tped and .bed from \texttt{plink} \citep{purcell07a}, VCF \citep{danecek11a}, and HapMap \citep{hapmap03}), as well as a generic text and binary file. Reading and writing in these formats is supported, as well as limited manipulation (see the reference manual for details). Format conversion is not supported at present. Random number generators and population indexing facilities are also available. The library is constructed using hierarchical classes and is built with extensibility in mind. File manipulations are implemented to reduce random-access memory (RAM) use, without unduly reducing execution speed. The trade-offs were tested on a laptop with a solid state drive (SSD) and 16 gigabytes of RAM. Performance may differ on other system types. 

In addition to the software, a directory with example SNP files is provided in the distribution for testing purposes. The project GitHub page (\url{https://github.com/tonymugen/sampleSNPs/}) provides a mechanism for users to report problems. Detailed library interface documentation is available at \url{https://tonymugen.github.io/sampleSNPs/}.

\paragraph*{Sampling variants.}
I used the \texttt{libsampFiles} library to write a stand-alone program, \texttt{sampleSNPs}, that subsamples variant files. All formats mentioned above are supported. The program runs via command line using standard Unix-style flags to pass execution parameters. The README file included with the project and available on the GitHub documentation page has detailed instructions. Sampled SNPs are saved into a file in the same format as the original. Auxiliary files, if present (e.g., .fam and .bim for .bed), are modified or copied as appropriate.
\paragraph*{Linkage disequilibrium among sampled loci.}
As an example of an application of locus sampling, I implemented a stand-alone program that estimates genome-wide LD decay with between-locus distance. A full accounting of this relationship would require the calculation of linkage disequilibrium statistics for all $N_p = n(n-1)/2$ pairs of loci, where $n$ is the number of genotyped variants. This task quickly becomes unmanageable as the number of genotypes in the data set grows. One solution, implemented in \texttt{plink} \citep{purcell07a}, is to calculate LD only among loci falling within a neighborhood window on a chromosome. A complementary approach: implemented here, is to sample $2N_s$ ($N_s$ is the desired number of sampled pairs) loci using Vitter's method and calculate LD between consecutive pairs. Justification for this approach is provided in Appendix A. Once a pair of loci is picked, \texttt{sampleLD} calculates two linkage disequilibrium statistics: $r^2$ and $D^{\prime}$ \citep{lewontin64}. Missing data are removed (only individuals successfully genotyped at both loci are considered). If there are not enough genotypes to produce a meaningful result, ``-9'' is reported. If a file with a population index is provided, the program will calculate LD statistics within each population and report them separately.

Unlike \texttt{sampleSNPs}, \texttt{sampleLD} currently only supports \texttt{plink} .bed files as input. The auxiliary .bim and .fam files are also required.  A detailed description of input file requirements, command line flags, and output format are in the README file included with the project and on the documentation page.

\subsection*{Test data}
Execution timing was performed with SNP files extracted from the \textit{Drosophila} genome nexus \citep{lack16a}. I used the Zambia and France populations from that data set. LD measurements were performed on cultivated rice (\textit{Oryza sativa}) genotypes \citep{mccouch16a}. I extracted a random sample of 100 \textit{indica} (IND) and 100 tropical \textit{japonica} accessions, and filtered out loci with minor allele counts less than two. I estimated the smoothed relationships between LD and distance, with their confidence intervals, using the \texttt{ggplot2} R package \citep{wickham09a}.

\section*{Results and Discussion}
After an extensive search I was unable to find existing software that performs uniform sampling of loci from files without replacement while preserving their order. The widely-used command-line tool \texttt{plink} \citep{purcell07a} does have a function (accessible via the \verb|--thin| flag) that extracts a random sample of SNPs while preserving order. However, the program simply examines each locus and includes it with the specified probability. Thus, the resulting sample varies in size (Supplemental Fig. S\ref{sfig:plinkSfig}).
\subsection*{Ordered sampling of loci}
Given that no other software appears to be available, I set out to implement a light-weight solution that quickly generates ordered samples without replacement even from very large data sets. The simplest idea is to examine every variant record in turn and decide, given the current number of loci picked, the number remaining in the input file, and the total to be sampled, whether to pick the current one \citep{fan62a}. The selected records are read into memory and saved to the output file. While this solution is obvious and easy to implement, it requires an examination and a (pseudo)random number sampling step for each line in the input file. 

An alternative approach has been proposed by Vitter \citep{vitter84a,vitter87a}. The idea is to decide how many loci to skip, given the current number already picked and remaining to be examined.  \citet{vitter87a} demostrated that this approach (Vitter's Method D) is faster than the simple line-wise decision-making outlined above (referred to as Method S). However, the tests were performed 30 years ago. The files were stored on tape, and random number generation was computationally expensive. Therefore, I implemented both Method S and Method D in C++, using comparable language facilities (see Methods and Supplemental files for details), and tested them in a number of scenarios to determine which scheme is preferable on modern-day computers.

Several variables can influence algorithm execution speed. Random (or pseudorandom) number generation is used extensively to generate samples from required distributions. However, code profiling (not shown) revealed that at least the hardware RNG I chose for this implementation (see Methods for details) is never the rate-limiting step, even when files are stored on a fast solid state drive. Rather, it is file input/output that takes most time during a given execution cycle. There are two parameters to consider when we investigate file read/write timing. One, storage can be either on a solid-state (SSD) or a spinning drive (HDD). The former is generally faster and allows for random file access. Second, the files can be either in a binary format (I use the popular and highly-compressed \texttt{plink} \texttt{.bed}), or in plain text. The important difference is that lines in files with text records are in general variable in length. Thus, if we want to skip several records we have no choice but to read each row in turn and discard unwanted loci. In contrast, binary formats use fixed-size fields, leading to uniform row sizes. It is then trivial to compute the number of bytes to  skip without reading before arriving at the desired record. 

\begin{figure}[htbp]
\renewcommand{\familydefault}{\sfdefault}\normalfont
\centering
	\includegraphics[width=\linewidth]{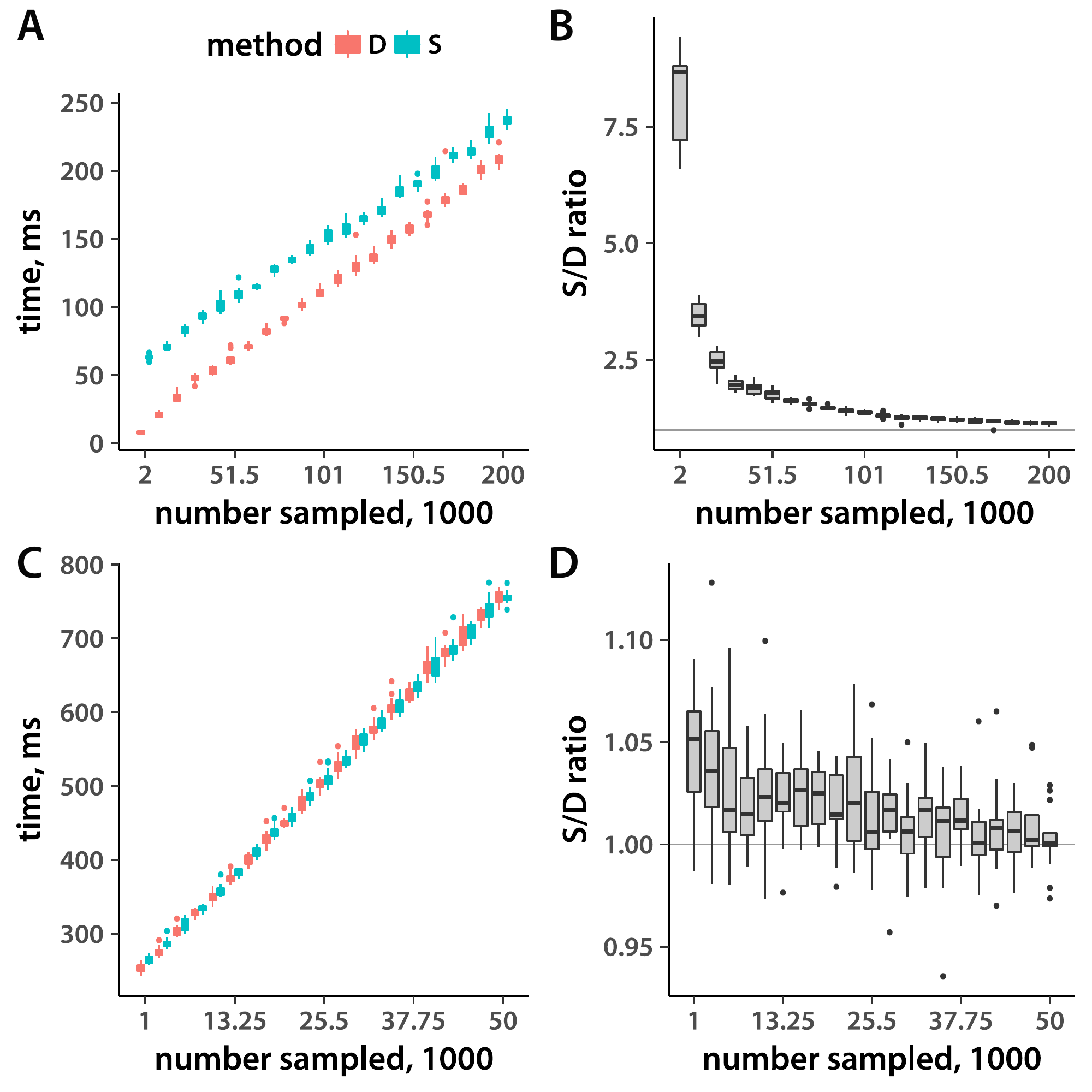}
	\caption{Execution timing with samples of varying size. (A) Execution time ($y$-axis) as a function of the number of samples picked from the file ($x$-axis). The total number of SNPs is held constant. The input file is in a binary (.bed) format. (B) The ratio of the Method S to Method D execution time as a function of the number of samples taken from a binary file (the timing data taken from panel A). (C) Arranged the same as panel A, but the input file is in a text (.tped) format. (D) The same as B, but using the .tped format. Distributions are derived from 15 replicate runs.}%
\label{fig:sampFig}
\end{figure}

Given that I am interested in creating a tool that can be used on personal workstations and laptops, and since solid state drives have become the standard choice, I focus on execution timing on a laptop (mid-2015 15-inch MacBook Pro) with an SSD. However, I also replicated the results on a 2014 Mac Mini with an HDD with essentially the same results (see Supplemental Figures \ref{sfig:hddSampFig} and \ref{sfig:hddConstFig}). I first held the input file size constant and varied the number of loci sampled. As shown in Fig.\ \ref{fig:sampFig}, time taken by both Method D and Method S grows approximately linearly with the number of loci sampled. This is the case for both binary and text files. Method D (Vitter's skip-over algorithm) outperforms the simpler Method S several-fold when the number of records is much smaller than the total in a binary file. This is not surprising, given that in this case Method S examines and discards many more loci for each one picked. As expected, the difference largely disappears when we sample from a text file. This is because both methods have to at least read and discard file lines one at a time. Interestingly, I obtained similar results on an HDD (Supplementary Fig.\ \ref{sfig:hddConstFig} and \ref{sfig:hddSampFig}), even though a spinning drive should not allow the same level of random access as an SSD. It is notable that in every case working with a binary file is about an order of magnitude faster, even though I am sampling more loci from a bigger file (500,000 loci in the binary \textit{vs} 100,000 in the text file). Finally, although the performance benefit of Method D is not always dramatic, it never underperforms Method S. The relatively small amount of extra time taken by Method S likely reflects the additional operations that are necessary to decide whether to include a given record. 

\begin{figure}[htbp]
\renewcommand{\familydefault}{\sfdefault}\normalfont
\centering
	\includegraphics[width=\linewidth]{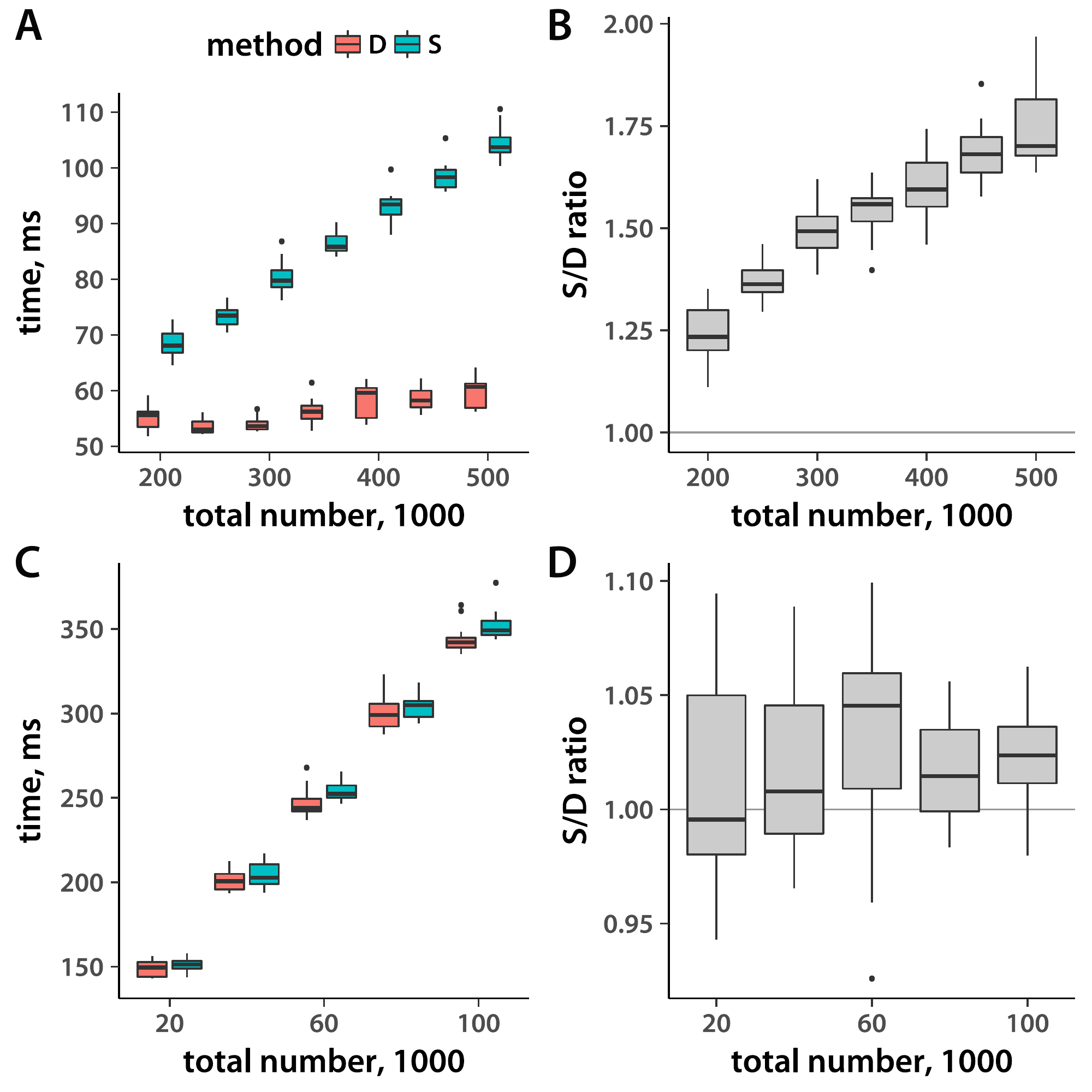}
	\caption{Execution timing and total record number. (A) Execution time ($y$-axis) of sampling 50,000 SNPs as a function of the total number of records in the (binary .bed) file. (B) The ratio of Method S to Method D timing, derived from the data in panel A. (C) The same as (A), but for sampling 10,000 loci from a text .tped file. (D) Method S to Method D execution time ratio for the data from panel C. Distributions reflect 15 replicate runs.}%
\label{fig:constFig}
\end{figure}

Given that Vitter's method decides ahead of time how many loci to skip before sampling, I would expect that it should be relatively insensitive to the total size of the input data set. Indeed, this is the case for binary file sampling (Fig.\ \ref{fig:constFig}, panels A and B). Increasing input file size 2.5-fold results in no measurable rise in execution time. Method S execution time, and that of both methods on a text file (Fig.\ \ref{fig:constFig}, C and D), grows approximately linearly with input size. Again, Method D is always at least as fast as Method S.

Given that Method S never consistently outperforms Vitter's Method D, I included only the latter in my implementation. While I do include the facility to read various text variant file formats, it is clear that using the \texttt{.bed} binary files, ideally on an SSD, results in optimal performance.

\subsection*{Linkage disequilibrium distributions}

Estimating rates of LD decay with distance on a chromosome are necessary, for example, in genome-wide association studies where such rates determine peak resolution. Because calculating LD between all pairs of loci is infeasible and unnecessary, a typical approach is to estimate linkage statistics in sliding windows. This technique is employed in \texttt{plink}. I implemented an alternative approach, picking loci according to Vitter's algorithm (see Methods for details) and then calculating LD statistics between consecutive variants in the sample. To test my implementation, I used a data set of 638,699 SNPs from the rice high-density array \citep[see Methods for details]{mccouch16a}. I first ran \texttt{plink} to calculate $r^2$ and $D^{\prime}$ between loci no more than 500 kb or 20 SNPs apart. This yields more than 12 million locus pairs, stored in a 1.2 gigabyte file. The relationships between linkage disequilibrium and distance are depicted in Fig.\ \ref{sfig:plinkSfig}(A, B). As expected, precision of LD estimates between distant loci diminishes due to undersampling. I then analyzed the same data set using my approach, sampling 30,000 SNP pairs (the resulting file occupies a mere 1.4 megabyte). While the confidence intervals from these estimates are wider (Fig.\ \ref{sfig:plinkSfig}A, B), the pattern of LD decay is the same as that captured by the considerably larger sample set produced by \texttt{plink}. Thus, my light-weight approach may be the best option when great precision is not required and computational resources are limited. 

An extra feature of my \texttt{sampleLD} program, unavailable in \texttt{plink}, is the ability to make separate LD estimates for each population present in a set of individuals. I illustrate this possibility by estimating linkage disequilibrium in \textit{indica} and tropical \textit{japonica} rice varietal groups (Fig.\ \ref{sfig:plinkSfig}C, D). It is well established \citep{mccouch16a} that LD levels are lower in \textit{indica}. My analyses recapitulate this pattern.

The software described in this report enables users to quickly generate subsets of large SNP or CNV data sets, opening the door to numerous applications that constrain resources available for genetic data manipulation. This work exemplifies the kinds of approaches needed to speed discovery cycles and empower researchers lacking access to expensive hardware.

\section*{Web resources}
\begin{description}
\item[Project name:] sampleSNPs
\item[Project homepage:] \url{https://github.com/tonymugen/sampleSNPs/}
\item[Project documentation:] \url{https://tonymugen.github.io/sampleSNPs/}
\item[Operating systems:] Unix-like (Linux, BSD, Mac OS)
\item[Programming language:] C++
\item[Other requirements:] No dependencies other than the C++11 standard library
\item[License:] BSD three-part
\end{description}

\section*{Acknowledgements}
  The idea for this project was spurred by Kevin Lawler's blog post on Vitter's work (\url{https://getkerf.wordpress.com/2016/03/30/the-best-algorithm-no-one-knows-about/}). The data for implementation testing were obtained from the \textit{Drosophila} Genome Nexus (\url{http://www.johnpool.net/genomes.html}) and the Rice Diversity website (\url{https://ricediversity.org/data/index.cfm}).

\bibliographystyle{genetics}
\bibliography{tony}

\section*{Supplemental Data}
  \paragraph*{Supplemental file 1 -- Timing of sampling schemes}
	This is an archive of the directory that contains the R and C++ code, as well as data files, necessary to reproduce the algorithm timing analyses presented in this manuscript. Compilation and running instructions are included.
  \paragraph*{Supplemental file 2 -- LD analyses}
	This is an archive of the directory that contains the R code and data files necessary to reproduce the linkage disequilibrium results presented in this report.
  \paragraph*{Supplemental file 3 -- Software source code}
    This is an archive of the directory that contains the source code of software described in this paper. Compilation and testing instructions are included. An up-to-date version can be found on GitHub (\url{https://github.com/tonymugen/sampleSNPs/}).

\section*{Appendix A: LD sample scheme derivation}
For the sample to accurately reflect whole-genome values, the probability of picking each pair must be equal. Furthermore, pairs must be sampled without replacement. To derive such a scheme, I order the list of all possible locus pairs as they would appear in a pair-wise relationship matrix. Since LD measures are symmetric, we need only concern ourselves with the upper (or, equivalently, lower) triangle of this matrix. The first row of the upper triangle lists the pairings of the first locus on a chromosome with all $n - 1$ subsequent loci other than itself. Next row lists $n - 2$ relationships between the second variant and the rest, excluding itself and the first locus. The process continues until we reach the last locus, which has no additional pairs. Sampling from this list will yield the desired uniform representation of locus pairs. Each variant on the list of pairs is represented $n - i$ times in a row, where $i = 1, \dots, n-1$ is the index of the locus position. Thus, instead of going through the pairs list (which contains $N_p = n(n - 1)/2$ elements) we can use a two-step scheme. We start by picking the first locus in a pair by sampling variants with weights reflecting the length of their run in the pairs list. We would then randomly pick another variant from the remaining loci. Finally, we go back to the first SNP or CNV in the pair and use it to sample the jump length to the next locus according to a weighted algorithm and repeat the process until we have the desired number of pairs. The initial sampling weight for locus $i$ under this scheme is 

\begin{equation*}
w_i = \dfrac{p_i}{\sum_i p_i},
\end{equation*}
where $p_i$ is the probability of sampling locus $i$. Since, as mentioned above, each variant is represented $n - i$ times on the list,

\begin{equation*}
p_i = \dfrac{n - i}{N_p} = \dfrac{n - i}{n(n - 1)/2}
\end{equation*}
Since $p_i$ are probabilities, $\sum_i p_i = 1$. This leads to the expression for $w_i$:
\begin{align*}
w_i &=\quad \dfrac{p_i}{\sum_i p_i}\\
 &=\quad p_i\\
 &=\quad \dfrac{2(n - i)}{n(n - 1)}\\
 &=\quad \dfrac{2}{n-1} - \dfrac{2i}{n(n-1)}\\
 &\approx\quad \dfrac{2}{n} - \dfrac{2i}{n^2} \quad\text{when $n$ is large}
\end{align*}
Thus, the deviation from an equal-weight random sampling (with all $w_i = \tfrac{1}{n}$) depends solely on the value approximately $\tfrac{2i}{n^2}$, which is tiny for large data sets ($n \ge 100,000$) we typically encounter. 

According to the scheme presented above, once we have the first locus in a pair, we would then sample randomly from the loci further down on the chromosome to obtain the second variant for LD calculations. The next round would then require us to go back in the file to the first locus in the pair and continue with our scheme. This step is potentially computationally expensive, especially for text files with variable-width lines. To eliminate this complication, I further simplify the algorithm by instead using Vitter's method to sample $2N_s$ ($N_s$ is the desired number of sampled pairs) loci. I then calculate LD between consecutive pairs of variants. A slight correction is needed only when more than one chromosome is present in the data set. In such cases, locus pairs that are located on different chromosomes are discarded and the additional pairs are sampled to restore the total to the required value. The resulting scheme approximates true uniform sampling very well when data sets are large and sample sizes are relatively small (preliminary tests suggested that sample sizes as large as 1/3 the total number of SNPs still yield reasonable results).

\section*{Figures}

\begin{figure}[htbp]
\renewcommand{\familydefault}{\sfdefault}\normalfont
\centering
	\includegraphics[width=\linewidth]{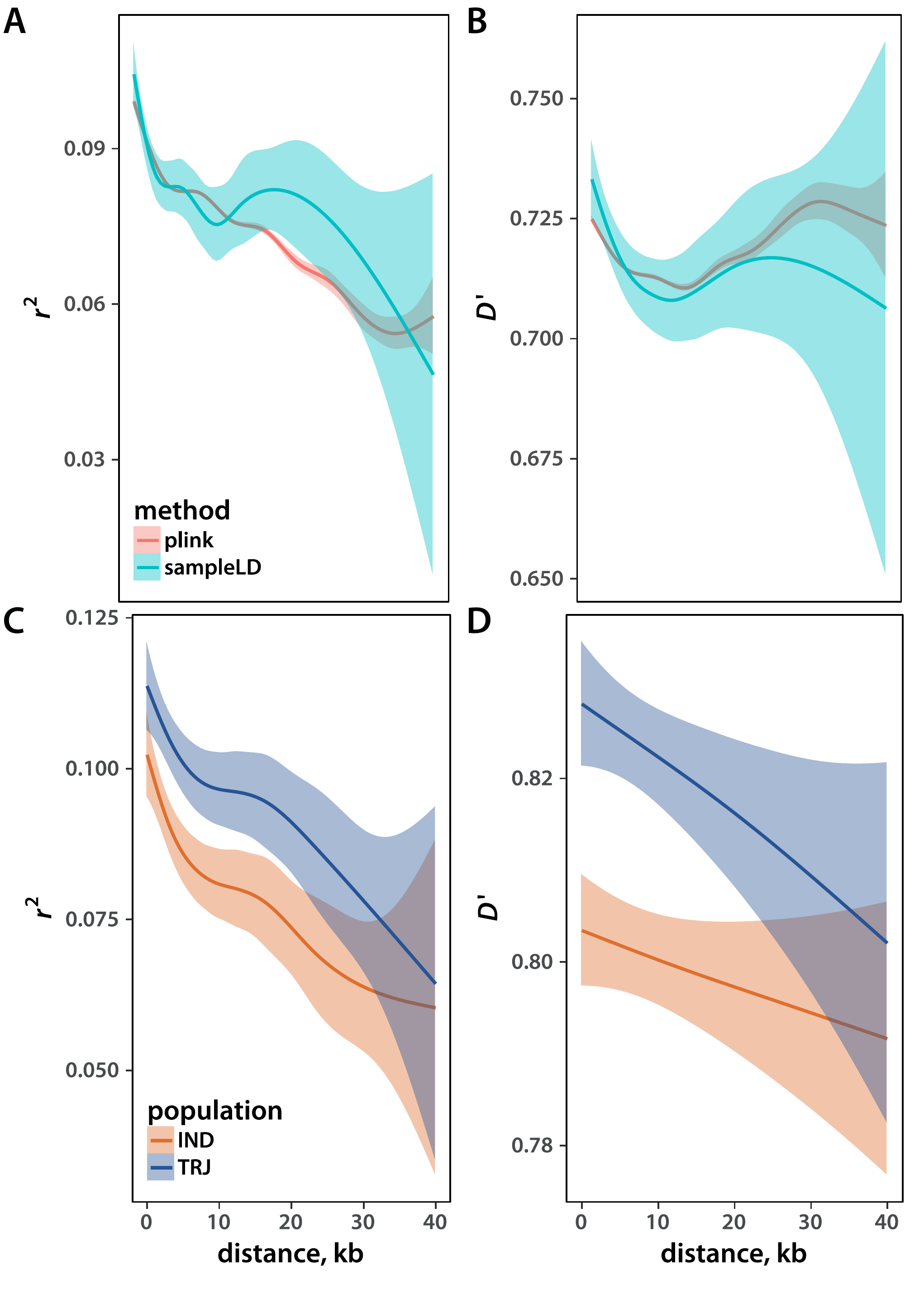}
	\caption{Linkage disequilibrium among sampled loci. The $r^2$ (A, C) and the $D^{\prime}$ (B, D) statistic as a function of distance between SNPs. Thick lines reflect the means and colored areas depict confidence intervals of a generalized additive smoothing function. Panels (A) and (B) compare the results from \texttt{sampleLD} and \texttt{plink}. Panels (C) and (D) depict estimates from \textit{O. sativa indica} (IND) and \textit{japonica} (JAP).}%
\label{fig:ldFig}
\end{figure}

\renewcommand{\figurename}{Supplemental Figure}
\setcounter{figure}{0}
\begin{figure}[htbp]
\renewcommand{\familydefault}{\sfdefault}\normalfont
\centering
	\includegraphics[width=\linewidth]{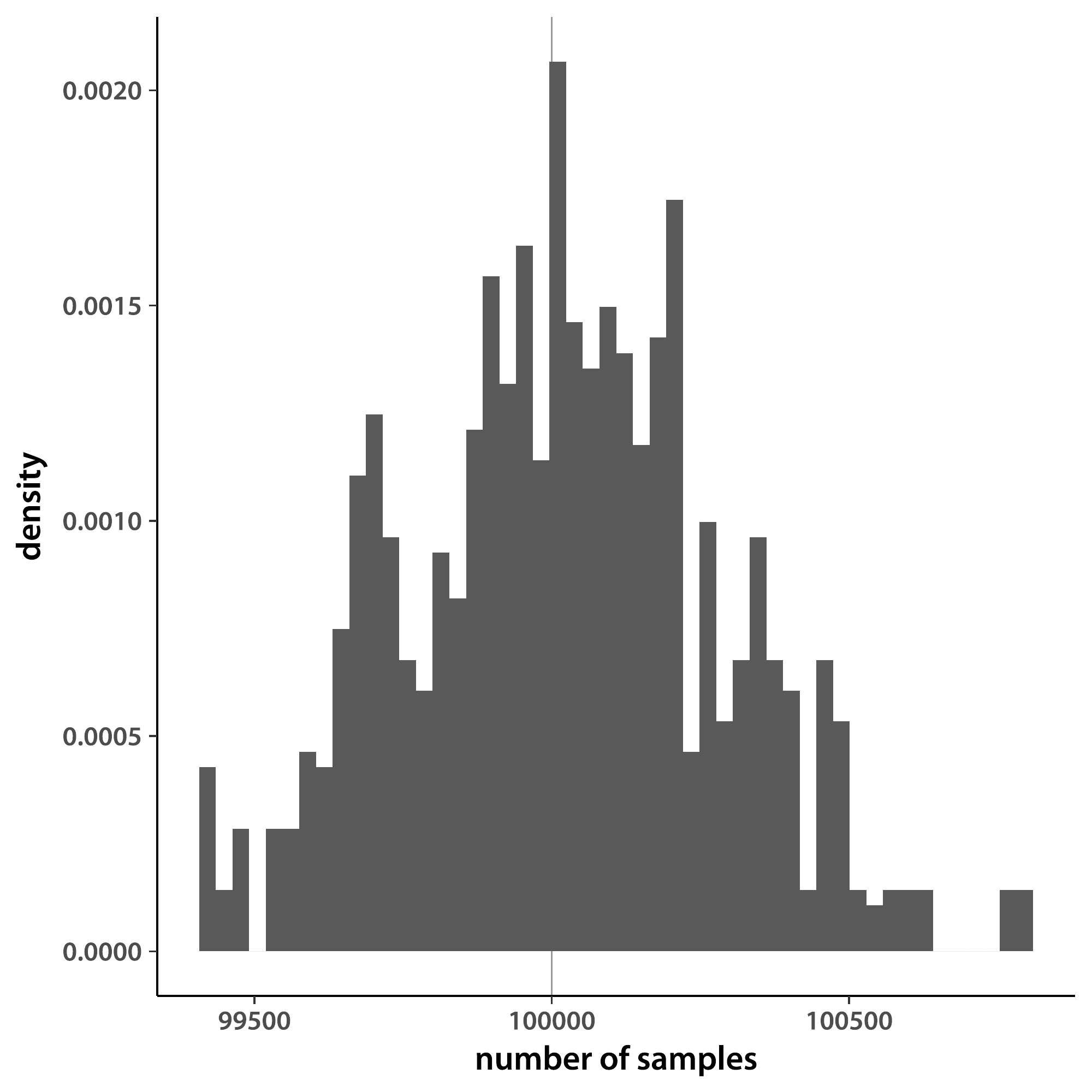}
	\caption{Distribution of sample sizes from \texttt{plink}. Number of samples returned by running \texttt{plink}, saving 20\% of the SNPs, 1000 times. Vertical line marks the expected number of samples.}%
\label{sfig:plinkSfig}
\end{figure}

\begin{figure}[htbp]
\renewcommand{\familydefault}{\sfdefault}\normalfont
\centering
	\includegraphics[width=\linewidth]{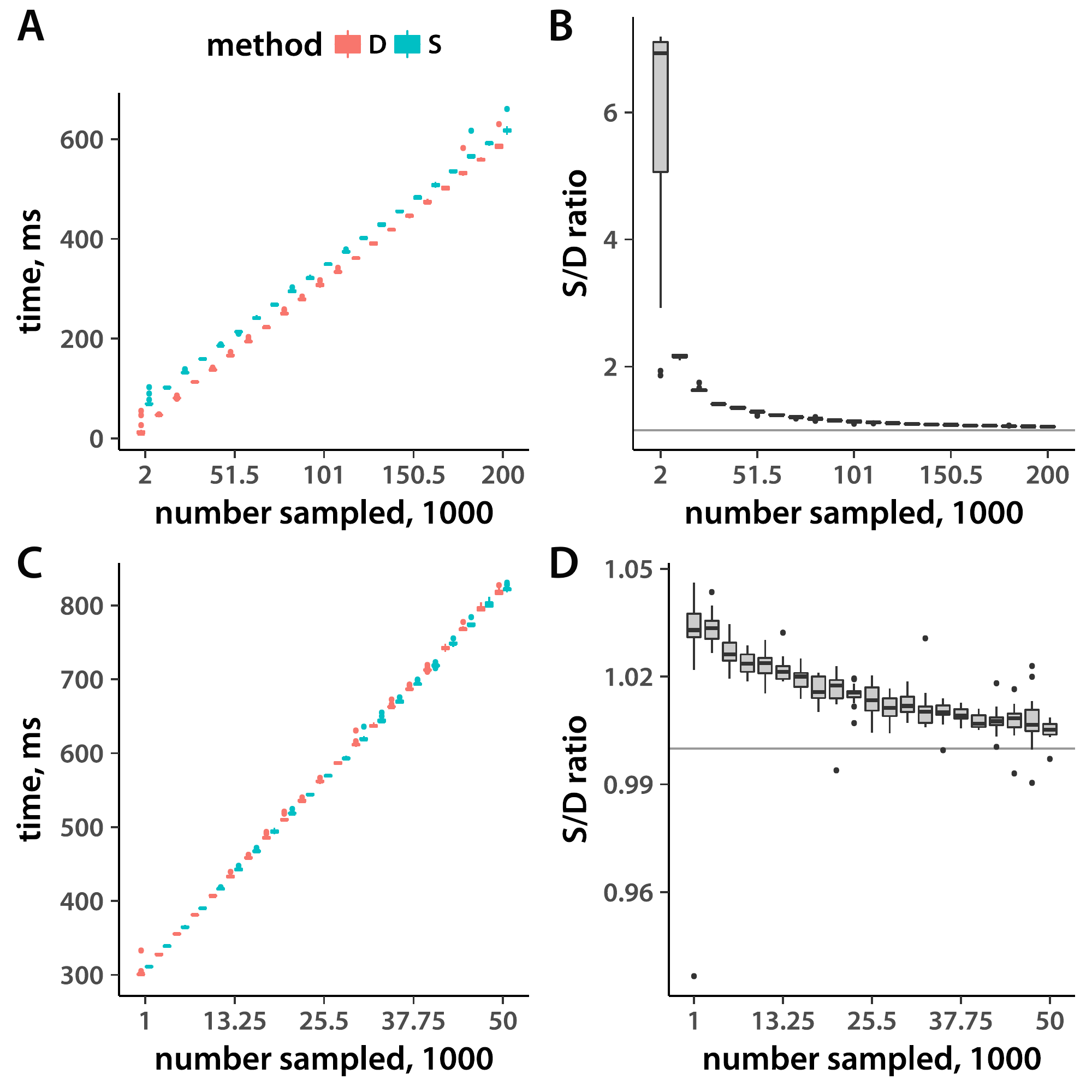}
	\caption{Execution timing with samples of varying size, HDD version. The figure is arranged the same as Fig.\ \ref{fig:sampFig}.}%
\label{sfig:hddSampFig}
\end{figure}

\begin{figure}[htbp]
\renewcommand{\familydefault}{\sfdefault}\normalfont
\centering
	\includegraphics[width=\linewidth]{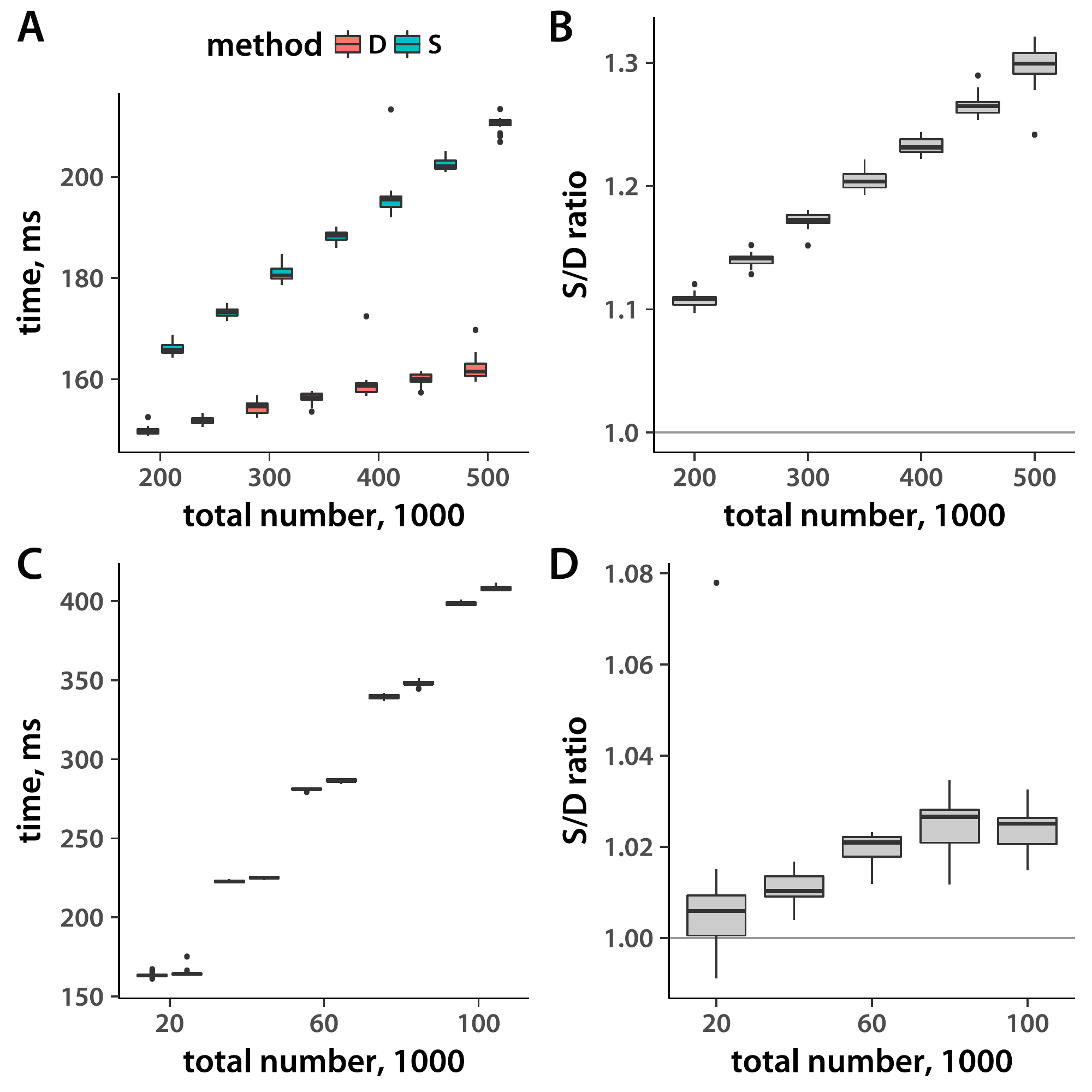}
	\caption{Execution timing and total record number, HDD version. The figure is arranged the same as Fig.\ \ref{fig:constFig}.}%
\label{sfig:hddConstFig}
\end{figure}

\end{document}